\documentclass[]{aa}  

\usepackage{graphicx}

\usepackage{txfonts}
\usepackage{hyperref}
\usepackage{orcidlink}

\usepackage{color}
\usepackage[dvipsnames]{xcolor}

\newcommand\nk{Necklace}
\newcommand\oiii{[O\,{\sc iii}]}
\newcommand\civ{[C\,{\sc iv}]}
\newcommand\ciii{[C\,{\sc iii}]}
\newcommand\cii{[C\,{\sc ii}]}
\newcommand\niii{[N\,{\sc iii}]}
\newcommand\nii{[N\,{\sc ii}]}
\newcommand\ha{H$\alpha$}
\newcommand\kms{\relax \ifmmode {\,\rm km\,s}^{-1}\else \,km\,s$^{-1}$\fi}
\newcommand\sm{M$_\odot$}

\newcommand\micron{$\mu$m}

\defcitealias{c11}{C11}

\begin{document}

   \title{Appraising the Necklace: A post-common-envelope carbon dwarf inside an apparently carbon-poor planetary nebula}
   \titlerunning{Appraising the Necklace}
   \authorrunning{D.\ Jones et al.}

   \subtitle{}

   \author{David Jones\orcidlink{0000-0003-3947-5946}\inst{1,2}
   \and
Romano L.~M. Corradi\orcidlink{0000-0002-7865-6056}\inst{1,3}
   \and
Gustavo A. Garc\'ia P\'erez\orcidlink{0009-0005-1276-0554}\inst{4}
   \and
Christophe Morisset\orcidlink{0000-0001-5801-6724}\inst{4, 5}
   \and
Jorge Garc\'ia-Rojas\orcidlink{0000-0002-6138-1869}\inst{1,2}
   \and
Laurence Sabin\orcidlink{0000-0003-0242-0044}\inst{4}
   \and
Bruce Balick\orcidlink{0000-0002-3139-3201}\inst{6}
    \and
Jacob Wise\orcidlink{0000-0003-0733-2916}\inst{7}
   \and
Antonio Mampaso\inst{1,2}
    \and
James Munday\orcidlink{0000-0002-1872-5398}\inst{8}
   \and
Pablo Rodr\'iguez-Gil\orcidlink{0000-0002-4717-5102}\inst{1,2}
   \and
Mar\'ia del Mar Rubio-D\'iez\orcidlink{0000-0003-4076-7313}\inst{9}
   \and
Miguel Santander-Garc\'ia\orcidlink{0000-0002-7338-0986}\inst{10}
   \and
Paulina Sowicka\orcidlink{0000-0002-6605-0268}\inst{1,2}
    \and
Alexander Csukai\orcidlink{0000-0003-0500-5997}\inst{11}
    \and
Todd C.\ Hillwig\orcidlink{0000-0002-0816-1090}\inst{12}
    \and
Andrea Henderson de la Fuente\inst{13,14}
    \and
Jacco H. Terwel\orcidlink{0000-0001-9834-3439}\inst{15}
          }

   \institute{Instituto de Astrof\'isica de Canarias, E-38205 La Laguna, Spain \email{djones@iac.es}
   \and
Departamento de Astrof\'isica, Universidad de La Laguna, E-38206 La Laguna, Spain
   \and
GRANTECAN, Cuesta de San Jos\'e s/n, E-38712, Bre\~na Baja, La Palma, Spain
   \and
Instituto de Astronom\'ia (IA), Universidad Nacional Aut\'onoma de M\'exico, Apdo. postal 106, C.P. 22800 Ensenada, Baja California, M\'exico
  \and
Instituto de Ciencias Físicas (ICF), Universidad Nacional Autónoma de México, Av. Universidad s/n, 62210 Cuernavaca, Morelos, México
   \and
Department of Astronomy, University of Washington, Seattle, WA 98195-1580, USA
    \and
Astrophysics Research Institute, Liverpool John Moores University, IC2, Liverpool Science Park, 146 Brownlow Hill, Liverpool L3 5RF, UK
   \and
Department of Physics, University of Warwick, Gibbet Hill Road, Coventry CV4 7AL, United Kingdom
   \and
Centro de Astrobiolog\'ia (CSIC/INTA), 28850 Torrej\'on de Ardoz, Madrid, Spain
   \and
Observatorio Astron\'omico Nacional (OAN-IGN), Alfonso XII, 3, 28014, Madrid, Spain
    \and
Jodrell Bank Centre for Astrophysics, Department of Physics and Astronomy, The University of Manchester, Oxford Road, Manchester M13 9PL, UK
    \and
Department of Physics and Astronomy, Valparaiso University, Valparaiso, IN 46383, USA
    \and
Department of Physics, University of Oxford, Parks Rd, Oxford OX1 3PU, United Kingdom
    \and
 Nordic Optical Telescope, Rambla Jos\'e Ana Fern\'andez P\'erez 7, 38711, Bre\~na Baja, Spain
     \and
School of Physics, Trinity College Dublin, The University of Dublin, College Green, Dublin 2, Dublin, Ireland
             }

   \date{Received October 2025; accepted January 2026}

\abstract
   {The Necklace nebula is a bipolar, post-common-envelope planetary nebula, the central star of which has been shown to have a dwarf carbon star companion.}
   {We aim to understand the origins of the Necklace and its dwarf carbon central star.}
   {We study the carbon abundance of the nebula through far ultraviolet spectroscopy obtained with the Hubble Space Telescope. Furthermore, through simultaneous modelling of multiband light and velocity curves, we attempt to constrain the parameters of the central star system.}
   {Puzzlingly, we find that the region of the inner nebula observed with the Hubble Space Telescope is seemingly not carbon-rich, at odds with the dwarf carbon star nature of the companion of the central star.  The initial mass of the nebular progenitor was likely very close to the limit to become carbon-rich, perhaps experiencing a very late thermal pulse.  The dwarf carbon star companion is found to be significantly inflated with respect to that expected for an isolated main sequence star of the same mass.}
   {The properties of the central binary are consistent with the progenitor having become carbon-rich and its companion having accreted a significant amount of that carbon-enriched material.  However, it is unclear how this evolutionary hypothesis can be reconciled with the inner nebula potentially being carbon poor.}
   \keywords{planetary nebulae: individual: PN~G054.2$-$03.4 -- binaries: close -- white dwarfs -- stars: AGB and post-AGB -- stars: chemically peculiar -- ISM: abundances 
               }

   \maketitle

\section{Introduction}

Dwarf Carbon (dC) stars are main-sequence stars which display molecular absorption bands of C$_2$, CN and CH in their spectra, even though they have yet to undergo the necessary nucleosynthesis or experience the third dredge up (TDU) required to have brought carbon-rich material to the stellar surface \citep{dahn77}.  The favoured hypothesis for the origins of such stars is that they have previously accreted carbon-rich material from a more evolved companion or its wind \citep[][]{green13}.  However, only a handful have been shown to be in spectroscopic binaries with close white dwarf (WD) companions \citep{dearborn86,margon18}, where the shortest period of these have almost certainly experienced a common-envelope (CE) phase \citep{roulston21,whitehouse21}.

The CE phase is perhaps the least well understood phase of binary stellar evolution \citep{ivanova13}, in spite of its critical importance in understanding the formation of \textit{almost all} close binaries with an evolved component (WD, black hole or neutron star). A CE is formed when runaway Roche lobe overflow in a binary system leads to the engulfment of the companion.  Subsequent tidal forces between the orbiting companion and donor's core lead to the deposition of orbital energy and angular momentum into the envelope, shrinking the orbit and lifting the CE either culminating in a merger or the formation of a close binary and the ejection of the CE \citep{paczynski76}. The binary star systems which survive this process, without merging, are thought to be the close-binary central stars of planetary nebulae (PNe), where the nebula itself comprises the ejected envelope while the central ionising source was once the degenerate core of the pre-CE donor star \citep{jones20ce}.

The \nk\ nebula \citep[IPHASXJ194359.5+170901, IRAS~19417+1701, PN~G054.2$-$03.4;][hereafter C11]{c11} is unique among post-CE PNe in that it has been shown to host a dC companion \citep{miszalski13} -- making it an invaluable system with which to probe the role of the CE phase in the formation of both PNe and dC stars.  The nebula's prominent ring morphology is thought to be due to strongly flattened deposition of the envelope, presumably in the orbital plane of the central binary \citep{hillwig16}.  The fainter, high velocity polar outflows, on the other hand, were found to have a larger kinematical age than the main body of the PN \citepalias{c11}. Therefore, they have been proposed to be formed by a collimated wind from an accretion disc around the secondary star {\it prior to} the CE phase \citepalias{c11}.  Evidence for a similar phase of pre-CE accretion is seen in other post-CE PNe, where the binary companions have been found to be inflated \citep{jones15,jones20png} and/or where the polar regions of the nebulae are kinematically older \citep{mitchel07}, however no other post-CE PN has been shown to host a dC central star.

In this paper, we present a detailed study of the Necklace in order to exploit its unique nature as a probe of the CE in the formation of highly aspherical PNe \citep{balick02,jones17} as well as the formation of dC stars \citep{green13}.  With this goal, we have modelled the light and radial velocity variations of the central binary in order to constrain the stellar and orbital parameters of the system.  We also probe the chemical and physical properties of the nebula via space-based narrowband imaging and ultraviolet (UV) spectroscopy.

\section{Observations}

\subsection{Optical spectroscopy}

Radial velocity measurements of the secondary star of the \nk\ nebula were obtained during several nights at the 3.5-m~APO telescope at Point Apache Observatory (New Mexico, USA) and at the 4.2-m~ William Herschel Telescope (WHT) at the Observatorio del Roque de los Muchachos (ORM, La Palma, Spain).

At APO, we used the DIS double-arm spectrograph.  In the blue, the B1200 grating with a slit width of 1.5'' provided a spectral reciprocal dispersion of 0.62~\AA\ per pixel, a resolution of 2~\AA, and spectral coverage from 3850 to 5050~\AA. In the red arm, grating R1200 provided a dispersion of 0.58~\AA~pix$^{-1}$, a resolution of 1.6~\AA, and wavelength coverage from 5650 to 6800~\AA.

At the WHT, we used the Intermediate-dispersion Spectrograph and Imaging System (ISIS).  In the blue arm, grating R600B was used, providing a dispersion of 0.44~\AA\ per pixel, a resolution of 1.6~\AA\ with the adopted slit width of 1$''$, and wavelength coverage from 3600 to 5050~\AA. In the red arm, grating R316R gave a dispersion of 0.92~\AA~pix$^{-1}$, and a resolution of 3.0~\AA\ from 5500 to 7300~\AA.

Typical integration times at each telescope were 30 or 40 minutes, which were repeated during the period of visibility of the target in the allocated time.  The journal of the spectroscopic measurements is presented in Table~\ref{T-logspec}.

\begin{table}
\caption{Journal of spectroscopic observations. For each night, the 
integration time and number of exposures obtained are indicated.}
\centering
\begin{tabular}{ll}       
\hline\hline\\[-7pt]                    
Date & Exposures \\
\hline\\[-7pt]              
{\it 4.2m WHT} & \\
2007-07-17 & 20 min $\times$3\tablefootmark{$\ast$}\\ 
2010-05-02 & 40 min         \\ 
2010-05-30 & 40 min         \\ 
2010-06-19 & 40 min         \\ 
2010-07-22 & 40 min         \\ 
2010-07-26 & 40 min         \\ 
2011-05-11 & 40 min         \\ 
2011-11-05 & 40 min $\times$2\\[2pt] 
{\it 3.5m APO} & \\
2012-06-20 & 30 min $\times$3\\ 
2012-06-21 & 30 min $\times$3\\ 
2012-06-23 & 40 min $\times$2\\ 
2012-06-24 & 40 min $\times$5\\ 
2012-06-25 & 30 min $\times$5\\ 
2012-07-13 & 40 min $\times$5\\ 
\hline\\[-5pt]
\end{tabular}
\tablefoot{
\tablefoottext{$\ast$}{Lower resolution spectra from \citetalias{c11}.}}
\label{T-logspec}
\end{table}

The APO and WHT spectra were reduced with the standard procedure using the {\it longslit} package of \textsc{iraf} V2.16\footnote{\textsc{iraf{}} is distributed by the National Optical Astronomy Observatory, which is operated by the Association of Universities for Research in Astronomy (AURA) under cooperative agreement with the National Science Foundation.}.

\subsection{Optical photometry}
\label{sec:phot}

In order to improve the orbital ephemeris determined by \citetalias{c11},
photometric monitoring was performed in the Sloan $i$-band on the night of 12 June 2013 at the 80cm~IAC80 telescope at the Observatorio del Teide (Tenerife, Spain), and in Sloan $g$, $r$ and $i$-bands sporadically from 21 August to 26 October 2023 with the Wide Field Camera (WFC) on the 2.5-m Isaac Newton Telescope (INT) at the ORM, with IO:O on the 2.0-m Liverpool Telescope (LT) on 14--16  April 2021,  with HiPERCAM on the 10.4-m Gran Telescopio Canarias (GTC) on 8 May 2021, and with the Alhambra Faint Object Spectrograph and Camera (ALFOSC) on the 2.56-m Nordic Optical Telescope (NOT) on 13 November 2024.  Typical individual integration times were 600--900-s for the IAC80, 90-s ($g$), 120-s ($r$) and 90-s ($i$) for the INT, LT and NOT, respectively, and 5--52s in all bands for the GTC. The new INT and NOT data, as well as the older INT, IAC80 and Mercator data presented in \citetalias{c11}, were reduced using standard routines of the astropy-affiliated python package {\tt ccdproc} \citep{ccdproc}, while those data from the IAC80, LT and GTC which were reduced with the respective instrument pipelines.  Differential photometry of the central stars was then performed against the field stars, IGAPSJ194358.44+170948.7 and IGAPSJ194357.83+170854.2 \citep[using the \textsc{photutils} package;][]{photutils}, before being placed on an apparent magnitude scale using the IGAPS catalogue photometry \citep{igaps}. The resulting photometric measurements are available at the CDS.

\subsection{HST optical imaging}

Hubble Space Telescope (HST) Wide Field Camera 3 (WFC3) images of the Necklace nebula were obtained by the Hubble Heritage Team under program 12675 on July 2011. Filters and exposure times are indicated in Table~\ref{T-hstimageslog}. Images were reduced by Max Mutchler at the Space Telescope Science Institute (STScI) and drizzled to half-pixel scale (0.02'' pix$^{-1}$) to enhance the spatial resolution.

\begin{table}
\caption{Journal of HST imaging}
\centering
\begin{tabular}{lcr}       
\hline\hline\\[-7pt]                    
Filter & $\lambda$/width [nm] & Exp. time [sec] \\
\hline\\[-7pt]              
F438W (B)        & 432.5/61.8   &  240          \\
F502N (\oiii)    & 501.0/6.5    &  2000         \\
F555W (V)        & 530.8/156.2  &  180          \\
F656N (\ha+\nii) & 656.1/1.8    &  2000         \\
F658N (\nii)     & 658.4/2.8    &  2000         \\
F814W (I)        & 802.4/153.6  &  120          \\
\hline\\[-5pt]
\end{tabular}
\label{T-hstimageslog}
\end{table}

\subsection{HST UV spectroscopy}
\label{sec:HST}
The \nk\ nebula was observed on July 2014 with the Cosmic Origins Spectrograph (COS) of the HST under program 13424 (PI: R.L.M.~Corradi). The circular 2.5'' diameter primary science aperture of COS was positioned 1.65'' north and 3.30'' west of the central star, intersecting the slit position of the optical spectra in \citetalias{c11} -- both of which are marked in Fig.~\ref{F-hstimages}. To save acquisition time, blind pointing to coordinates $\alpha=19^h43^m59.27^s$, $\delta=+17^\circ09'2.73''$ (J2000.0) was adopted, as the nebula is sufficiently uniform within 1 arcsec from the adopted target position.  The total exposure time on target was 2528.6 sec. Grating G140L was used with central wavelength setting at 110.5~nm: this provides a useful, calibrated spectrum from 112 to 215~nm.  The spectral reciprocal dispersion of this setting is 0.008~nm, but the effective spectral resolution is lowered to $\sim$0.7~\AA\ by the large aperture which is filled by gas emission from the \nk\ nebula. The pipeline calibrated spectrum was then rebinned by a factor of 18 pixels to improve the signal-to-noise per pixel while maintaining the spectral resolution (the spectral resolution element comprises roughly four post-binning pixels).

\section{Distance}

The Gaia parallax of the central star of the Necklace, $\omega=0.171\pm0.098$ mas \citep{chornay21}, leads to a distance of $\sim$5.4 kpc although with a relatively large uncertainty \citep[roughly $\pm$2 kpc;][]{bailer-jones21}.  This is generally consistent with the $\sim$5 kpc distance derived using the H$\alpha$ surface brightness-radius relationship of \citet{frew16} and the published integrated H$\alpha$ fluxes \citep[e.g.; \citetalias{c11} and][]{frew13}.  These distances lead to a height above the Galactic plane of $\sim$300 pc making it likely to be a member of the thin disk \citep{quireza07}, consistent with the borderline Peimbert type \textsc{i/ii} abundances and the small deviation of the radial velocity relative to that expected from the Galactic rotation curve  \citepalias[$|\Delta V| \sim$50~km~s$^{-1}$ assuming a circular orbit;][]{c11}.

\citet{dharmawardena21} derive a significantly larger distance ($>$8~kpc) using the extinction-distance mapping of \citet{sale14}. However, their distance is based on a visual extinction, $A_V$, of 2.41$\pm$0.19~mag \citep[measured by comparing the essentially unextincted radio free-free flux of the nebula to the optical H$\alpha$ flux from the IPHAS survey;][]{drew05} which is significantly larger than the value derived from the spectroscopic Balmer line ratios \citepalias[$A_V=1.19\pm0.12$~mag;][]{c11}.  Furthermore, if there is significant internal extinction, this would artificially increase the derived distance as it assumes only interstellar extinction. For the visual extinction of \citetalias{c11}, the extinction-distance mapping of \citet{sale14} implies a distance of roughly 2.7~kpc, appreciably closer than the other distance estimates outlined above. While the uncertainties are large, the Gaia distance of $\sim$5.4 kpc would appear to be the most reliable available.  At such a distance, the kinematical age of the nebular ring would be $\sim$6 kyr.

\section{The morphology of the inner nebula}

\begin{figure*}
\centering

\includegraphics[width=0.95\textwidth]{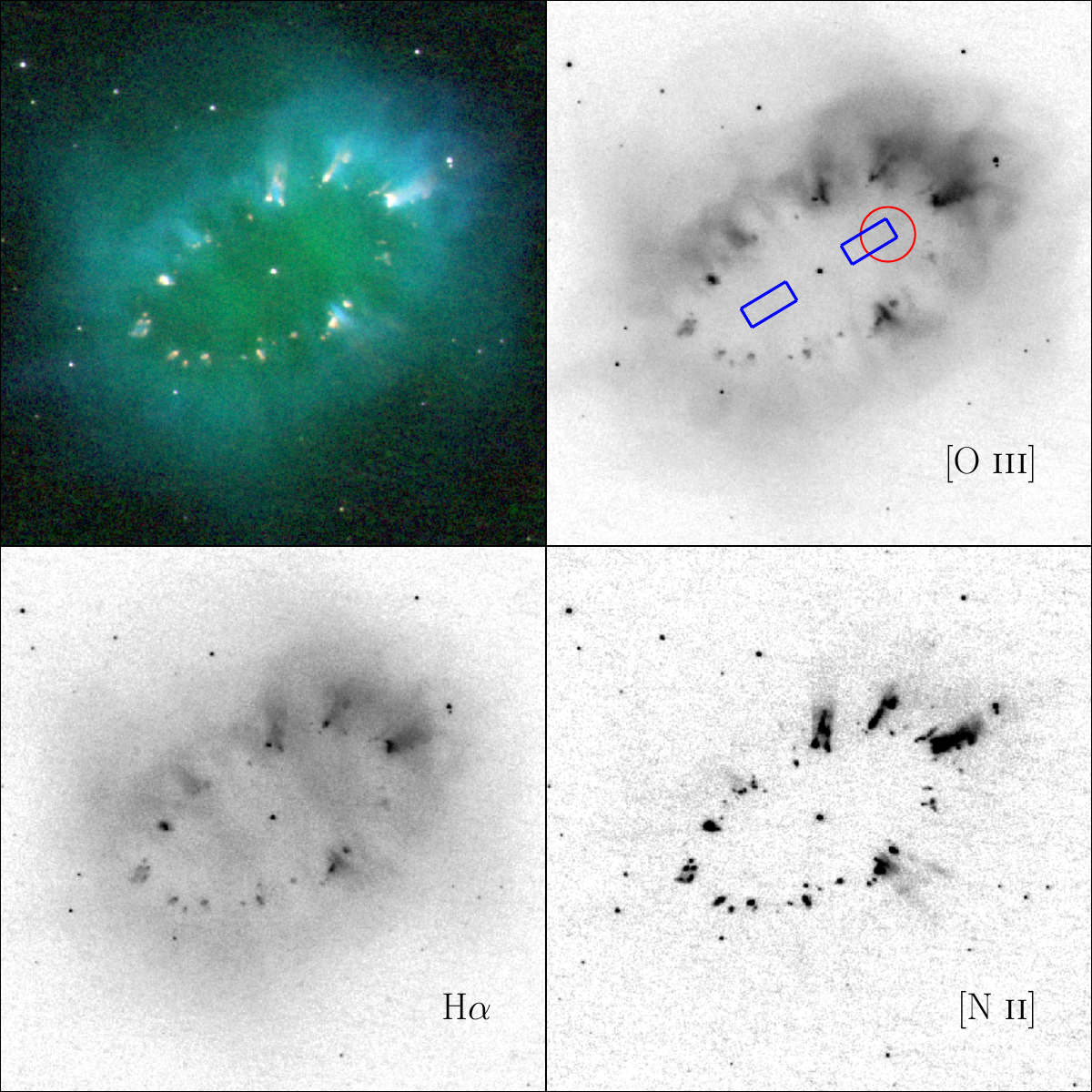}
\caption{HST images of the \nk. 
The field of view of each panel is 25$\times$25~arcsec$^2$. North is up and east is left.  The colour-composite image (upper left) comprises Red: \nii{}, Green: \ha{} and Blue: \oiii{}.  In the \oiii\ image, the position and size of the circular COS aperture is shown in red, while the ``inner nebula'' region from \citetalias{c11} is shown in blue.}
\label{F-hstimages}
\end{figure*}

The HST WFC3 images provide new insights into the remarkable ``necklace'' appearance of the nebula.  They are presented in Fig.~\ref{F-hstimages}. In the light of \oiii\ and \ha, the main body of the nebula and the environs of its knotty ring are clearly visible and, as described in \citetalias{c11}, define a main geometry that seems to be composed by a flattened, ``equatorial'' gas distribution, with fainter emission protruding in the ``polar'' directions.  The knots forming the necklace are also visible in both bands along with outward facing tails that are most clearly visible in the low-ionisation \nii\ line.  These tails can be due to photoionisation shadows or, perhaps more likely, ablation of the knots \citep{mellema98,steffen04,raga05,matsuura09,meaburn10}.  The colour-composite image presented in the upper-left panel of Fig.~\ref{F-hstimages} highlights the ionisation structure present in the knots, with the relative positions of \nii{} and \oiii{} being strongly indicative of UV ionisation (though shock ionisation cannot be ruled out).

Compared to the ground-based images of \citetalias{c11}, most knots are, albeit marginally, resolved. About fifty knots can be counted, some of them organized in groups, and without constant spacing (although the principal clusters of knots seemingly all have diametrically opposed counterparts).  Their sizes range from close to the HST point spread function, 0.08 arcsec (corresponding to 430 au at a distance of 5.4 kpc), up to 0.5 arcsec (2100 au), but note that the larger knots could be composed of smaller, unresolved complexes, and are less centrally condensed consistent with being at an advanced stage of evaporation. To roughly estimate their mass, one can assume approximately 50 knots with a typical diameter of 500 au and an electron density, $n_\mathrm{e}$, of 800 cm$^{-3}$ \citepalias[based on the NW knot density measured by][]{c11}, leading to an estimate of 1$\times$10$^{-5}$ M$_\odot$ for the total mass in the knots -- a factor of $\sim$10$^3$ lower than the total ionised mass estimated by \citetalias{c11}.

\section{Updated light curve ephemeris}
\label{sec:ephem}

The spectra in Table~\ref{T-logspec} are taken many hundred orbits after the original photometric data of \citetalias{c11}. This could result in significant uncertainty in the extrapolation of the orbital phase to the dates of the spectroscopic observations. For this reason, the light curve ephemeris was revised using all the available photometric data (as described in Section~\ref{sec:phot}).

An analysis of the full $i$-band dataset results in the following refined ephemeris\footnote{The newly refined orbital period is slightly longer than the period derived in \citetalias{c11}, but still within twice their quoted uncertainty.  The times of reference minimum show poorer agreement, with the uncertainty quoted in \citetalias{c11} seemingly an order of magnitude too low.} for the Heliocentric Julian Date of the photometric minimum:
\begin{equation}
\mathrm{HJD}_\mathrm{min}=245\,5075.2080(8) +  1.161393(1) E
\end{equation}
where $E$ is an integer representing the number of orbital cycles since the reference time of minimum.

\begin{figure}
\centering
\includegraphics[width=\columnwidth]{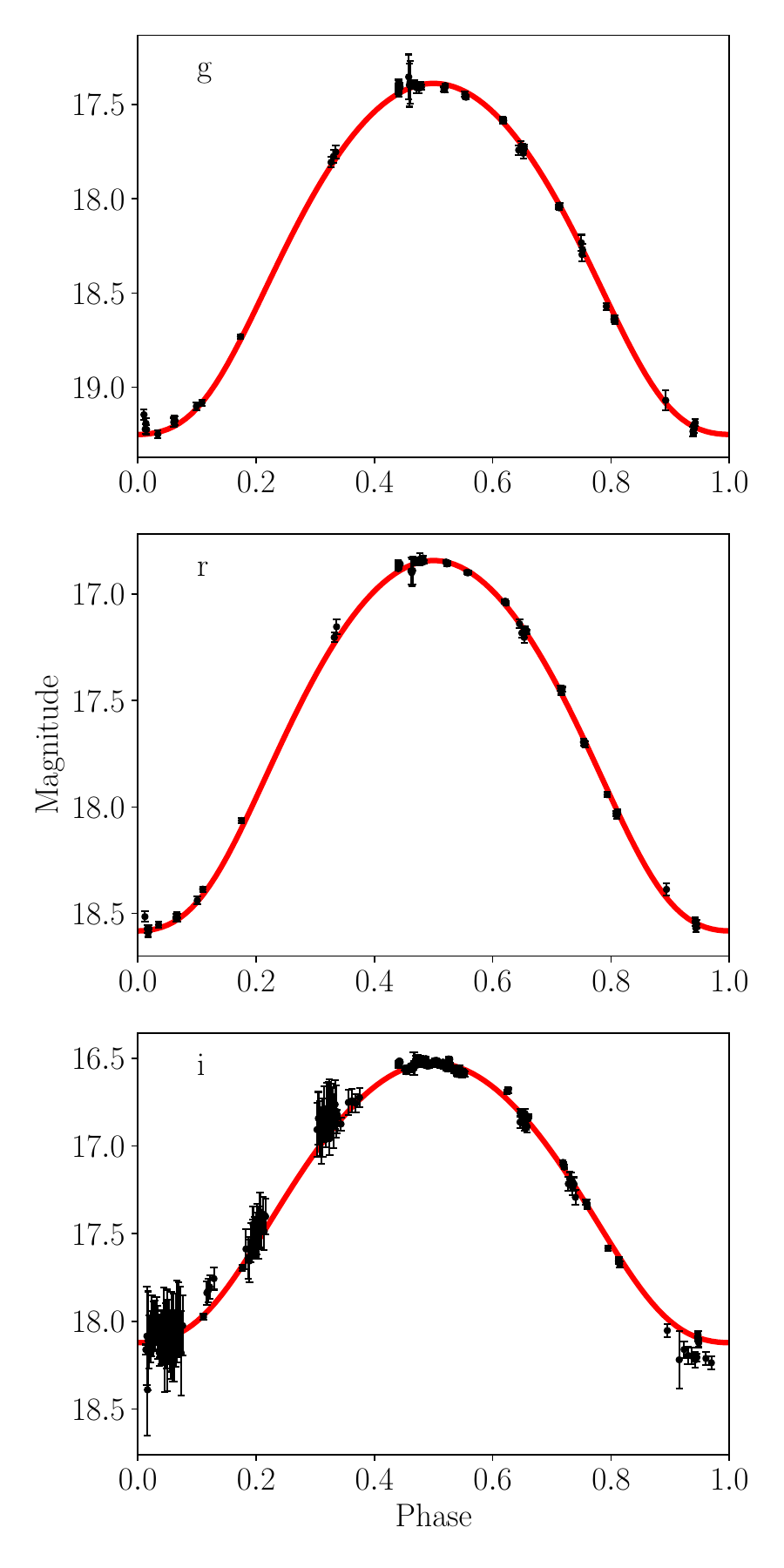}
\caption{Phase-folded light curves along with the associated \textsc{phoebe2} fits.}
 \label{F-phoebe}
\end{figure}

The photometric data in all three bands ($g$-, $r$- and $i$-band) are shown folded on this ephemeris in Fig.~\ref{F-phoebe}.  The light curves all display roughly the same sinusoidal morphology, typical of non-eclipsing irradiated binaries \citep[e.g.][]{exter03}.  Somewhat atypically, the amplitude of variability appears to decrease with the effective wavelength of the filter \citep[cf.][for example]{munday20}.  There also appears to be some additional scatter beyond the level of the purely statistical uncertainties (perhaps most notably in the $i$-band data from the IAC80 between phases 0.9--1.0), which may potentially be due to under/over-subtraction of the nebular background, differing filter/instrument responses or even intrinsic variability on top of the irradiation effect.

\section{Radial velocity curve}
\label{sec:rvs}

\begin{table}
\caption{Stellar emission lines used in the radial velocity curve, their  adopted rest wavelengths (in air).}
\centering
\begin{tabular}{lll}       
\hline\hline\\[-7pt]                
Ion & \multicolumn{1}{c}{$\lambda_\mathrm{rest}$}  & Notes \\
    & \multicolumn{1}{c}{[\AA]}           &           \\
\hline\\[-5pt]                     
\cii            & 6462.04, 7231.34, 7236.80 & blend\\ 
\ciii\ ``blue'' & 4647.43, 4650.65& uncertain $\lambda_\mathrm{rest}$ \\ 
\ciii\ ``red''  & 5695.92, 6744.39 & \\ 
\civ            & 4658.33, 5801.34, 5811.97  & \\
\niii           & 4640.64 & \\
\hline\\
\end{tabular}
\tablefoot{Wavelengths from
\href{http://www.pa.uky.edu/~peter/atomic/}{http://www.pa.uky.edu/~peter/atomic/}.}
\label{T-wave}
\end{table}

\begin{figure}
\centering
\includegraphics[width=\columnwidth]{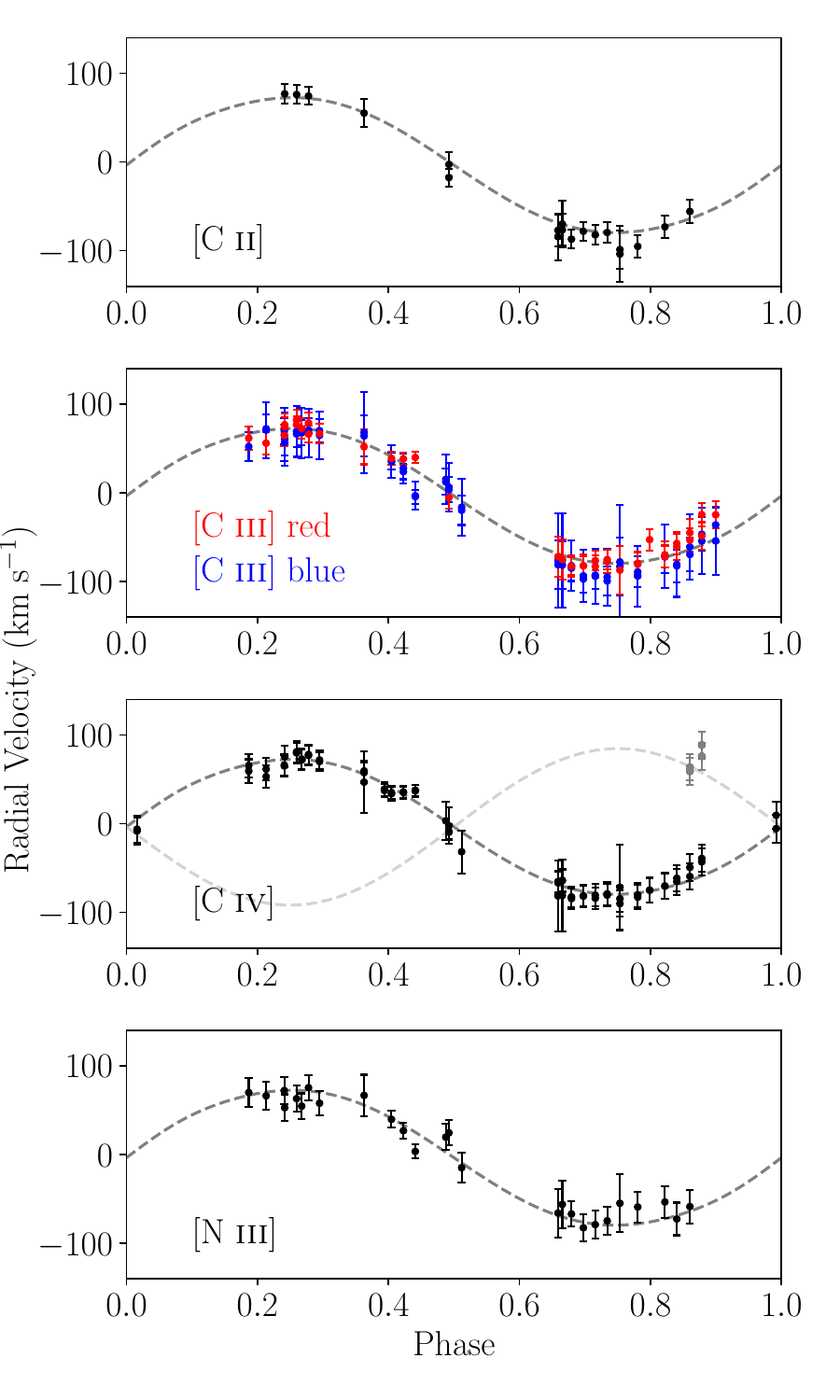}
\caption{Radial velocity curve of the secondary star of the \nk\ for different ions listed in Table~\ref{T-wave} The emission lines used are listed in Table~\ref{T-wave}. The underlaid dashed lines are the \textsc{phoebe2} model RVs ($K_2\approx$77.5~km\,s$^{-1}$). The grey points on the [C~\textsc{iv}] plot are tentatively associated with the hot primary (see text), while the lighter grey curve on that same subplot are the \textsc{phoebe2} model RVs for that component.}
 \label{F-rv}
\end{figure}

First, the spectrum of the central star was extracted from the long-slit spectra using the iraf {\it apall} task. The contribution of the sky and of the extended PN, which does not show emission lines coincident with the stellar emission lines, were not removed because they were used to adjust the zeropoint of the wavelength calibration at each epoch. Then, radial velocities (RVs) were measured by multiple Gaussian fitting of the strongest stellar emission lines. These lines are assumed to be produced by Bowen fluorescence on the hemisphere of the secondary star irradiated by the hot post-AGB primary \citepalias{c11}. The emission lines considered for the radial velocity (RV) measurements, and their adopted wavelengths are listed in Table~\ref{T-wave} (see also Fig.6 of \citetalias{c11}).

The resulting RV curves, phased with the light curve ephemeris, are presented in Figure~\ref{F-rv}. Individual errors depend on several factors, mainly the signal-to-noise (S/N) ratio of the emission features considered (which varies with the orbital phase) and blending of nearby lines or multiplets of the same ion.
Data are only presented for ions and transitions with good signal-to-noise and not affected by severe blending. Other emission lines show consistent behaviour but with larger scatter.

The RV curves of all species present with very similar amplitudes, indicating that in spite of their differing ionisation stages they originate from roughly the same region of the irradiated face of the companion (i.e.\ the centre of light for each line reflects roughly the same distance from the secondary's barycenter). We conservatively estimate the semi-amplitude, $K_2$, from all the ions to be approximately of 77.5~\kms, noting that this is likely a lower limit given that the irradiated emission lines do not reflect the centre-of-mass velocity \citep{jones20png}. This corresponds to a mass function $f(M_1)=\frac{{M_1}^3\sin^3i}{(M_1+M_2)^2}\gtrsim$5.6$\times10^{-2}$~\sm. Assuming that the orbit is oriented at an inclination $i=59$\degr{} as the nebular ring \citep{hillwig16}, this implies a mass ratio $q=M_2/M_1>1$ for all but the lowest mass remnants (i.e.\ for $M_1\gtrsim0.35$~\sm{}, with some uncertainty on the exact lower limit due to the estimated $K_2$ being a lower limit as discussed earlier).

Note that in the spectra from both \civ\ lines at 5801 and 5812~\AA, obtained near phase 0.9 (close to the inferior conjunction of the companion, plotted in grey in Fig.~\ref{F-rv}), a second velocity component appears, which we tentatively associate with the hot post-AGB star.  If this is indeed the case, then the points on either side of phase 0 may also originate from the post-AGB primary.

\section{Nebular C/O ratio}
\label{sec:co}

\begin{table}
\centering
\caption{COS-HST dereddened line fluxes }
\begin{tabular}{lcc}       
\hline\hline\\ [-7pt]                    
Line & $\lambda$ (\AA) & I($\lambda$)/I(H$\beta$)  \\
\hline\\[-7pt]              
N~{\sc v} & 1240$^\ast$ & 0.210$^{+0.059}_{-0.046}$ \\ [3pt]
Si~{\sc iv} & 1394 & 0.216$^{+0.046}_{-0.038}$ \\ [3pt]
O~{\sc iv}]$^\ast$ & 1402 & 1.184$^{+0.247}_{-0.205}$ \\ [3pt]
N~{\sc iv}$^\ast$ & 1484 & 1.543$^{+0.290}_{-0.245}$ \\ [3pt]
C~{\sc iv}$^\ast$ & 1549 & 2.320$^{+0.413}_{-0.350}$ \\ [3pt]
He~{\sc ii} & 1640 & 7.765$^{+1.313}_{-1.123}$ \\ [3pt]
C~{\sc iii}]$^\ast$ & 1907 & 2.616$^{+0.479}_{-0.405}$ \\ [3pt]
\hline\\
\end{tabular}
\newline\noindent$^\ast$ Blend of several lines of the same ion.
\label{T-COS-lines}
\end{table}

\begin{figure*}
\centering
\includegraphics[width=\textwidth]{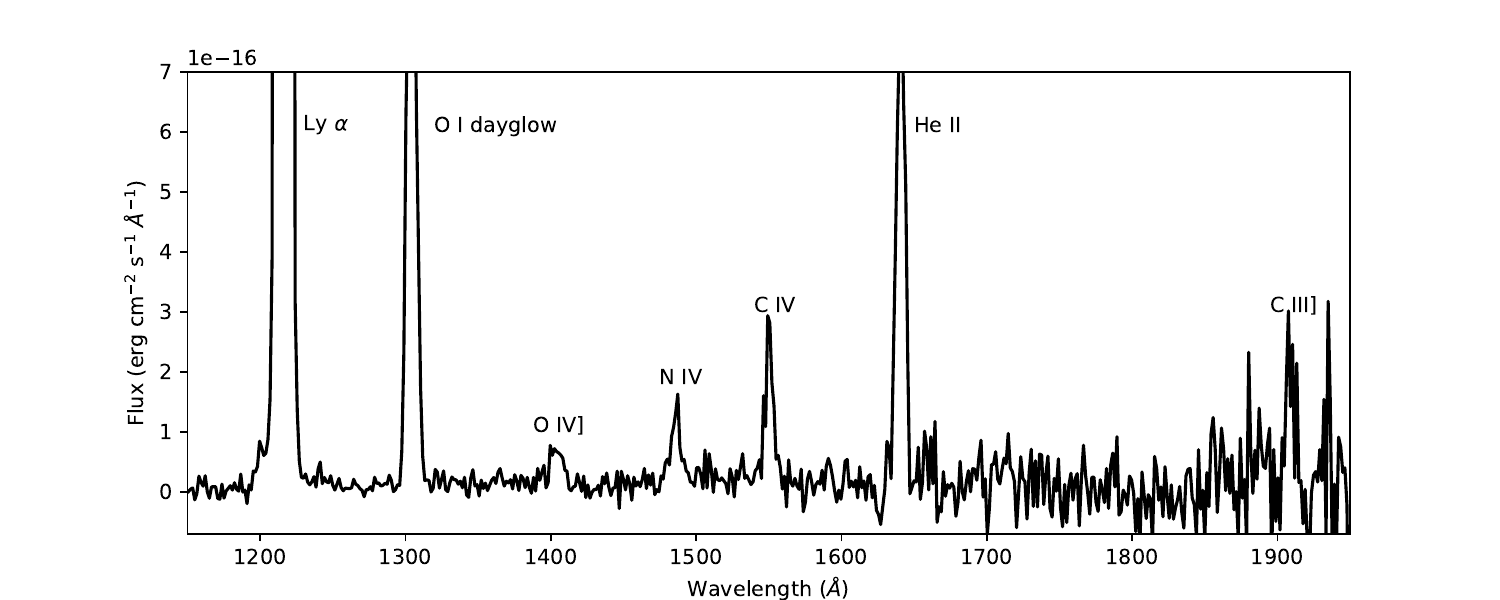}
\caption{COS-HST spectrum of the Necklace. The most remarkable features are labeled.}
 \label{F-COSspectrum}
\end{figure*}

\citetalias{c11} computed nebular abundances for two regions in the nebula: the inner nebula and a knot located northwest from deep optical spectra. One of the regions selected to extract the inner nebula spectra is almost position-coincident with the COS-HST pointing (see lower panel of their fig.~2 and top panel of our Fig.~\ref{F-hstimages}). We have taken advantage of such deep optical spectra and combine it with our deep COS UV spectra to try to shed some light on the chemical content of the inner nebula, particularly regarding the C/O ratio, which can be computed thanks to the detection of C~{\sc iii}] and [C~{\sc iv}] lines in the UV spectrum. In Fig.~\ref{F-COSspectrum} we show the smoothed COS spectrum including labels in the position of the most relevant detected emission lines.

To combine the UV (described in Sec.~\ref{sec:HST}) and optical spectra \citepalias[presented in][]{c11} for the determination of carbon abundances, a consistent normalization between both wavelength ranges was required. We adopted the optical data of \citetalias{c11} for the ``Inner nebula'', corresponding to a region comparable to the aperture covered by the HST-COS observations (see Fig.\ \ref{F-hstimages}). The normalization between UV and optical spectra was established using the theoretical He~{\sc ii} $\lambda1640/\lambda4686$ ratio. Taking the values derived for the inner nebula by \citetalias{c11}, we assume the observed intensity ratio, He~{\sc ii}~($4686$)/H$\beta = 111.6$, and the physical conditions ($T_{\mathrm{e}} = 14\,800 \pm 500$~K and $n_{\mathrm{e}} = 360 \pm 300$~cm$^{-3}$), and then derive the extinction-corrected ratio of $I(1640)/\mathrm{H}\beta = 776.22$ (in units of H$\beta = 100$) using {\sc pyneb} v1.1.18 \citep{luridianaetal15, morisset20}.  This cross-calibration implicitly accounts for the small aperture differences between the optical and UV data, assuring internal consistency between the UV and optical lines. Reddening correction was applied using the extinction law of \citet{fitzpatrick99} with $R_{V} = 3.1$ and $c(\mathrm{H}\beta) = 0.55 \pm 0.06$, again following \citetalias{c11}. Given that our line ratios are UV line fluxes relative to the optical, the  uncertainties on these ratios are completely dominated by uncertainty on the extinction. The final adopted line ratios and their uncertainties are shown in Table~\ref{T-COS-lines}.

\begin{table*}
\centering
\caption{Atomic data set used for abundance computations with {\sc pyneb}.}
\label{tab:atomic}
\begin{tabular}{lll}
\hline
Ion & Transition probabilities & Collision strengths \\
\hline
O$^{2+}$  & \citet{FroeseFischer:04} & \citet{Storey:14} \\
          & \citet{Storey:00} &  \\
O$^{3+}$   & \citet{Galavis:98} & \citet{Blum:92} \\
C$^{2+}$  & \citet{Glass:83} & \citet{Berrington:85} \\
          & \citet{Nussbaumer:78} &  \\
          & \citet{Wiese:96} &  \\
C$^{3+}$   & \citet{Wiese:96} & \citet{Aggarwal:04} \\
\hline
\end{tabular}
\end{table*}

{\sc pyneb} was then used to compute chemical abundances from both UV and optical spectra, and adopting the physical conditions derived by \citetalias{c11}. Uncertainties in the computed abundances are the result of the propagation of extinction and $T_{\rm e}$ uncertainties. The atomic data used are listed in Table~\ref{tab:atomic}.
The first estimate of the C/O abundance ratio is obtained directly from the sum of the available ionic abundances, i.e. $\mathrm{C/O} \simeq (\mathrm{C^{2+}} + \mathrm{C^{3+}})/(\mathrm{O^{+}} + \mathrm{O^{2+}} + \mathrm{O^{3+}}) = 0.158 ^{+0.315}_{-0.209}$ (but see below using ad-hoc ICFs). To compute the O$^{3+}$/H$^+$ ratio we considered the contribution of several O~{\sc iv}] lines at $\lambda\lambda$1397, 1400, 1401, 1404, and 1407 to the reported flux; similarly, C~{\sc iv} $\lambda\lambda$1548, and 1551, are considered in the C~{\sc iv} $\lambda$1549 reported flux. We did not attempt to correct the O~{\sc iv}] emission for possible contamination by Si~{\sc iv} $\lambda$1403, since the contribution is expected to be small (at most $\sim$18\%) given the weak flux measured in Si~{\sc iv} $\lambda$1394, which is always brighter than the Si~{\sc iv} $\lambda$1403 line under any opacity conditions.  
\subsection{Looking for adapted ICFs using 3MdB and Machine Learning method}

The C/O ratio computed above may be underestimated if a significant fraction of carbon is present as C$^{4+}$, which is plausible given the high degree of ionisation in the inner nebula (He$^{2+}$/He $\approx 0.90$) and the detection of [Ne~{\sc v}] lines, whose parent ion has a higher ionisation potential than those of C$^{4+}$. An alternative approach to simply summing the available ionic abundances is to use the ionisation correction factors (ICFs) of \citet{delgadoinglada14} and the optical data alone, which leads to lower oxygen abundances, $\mathrm{O/H} = 3.96 \times 10^{-4}$, and a higher carbon abundance, $\mathrm{C/H} = 3.54 \times 10^{-4}$, resulting in $\mathrm{C/O} \approx 0.89$ (with uncertainties that span C-rich values). However, this latter value should be considered highly uncertain, since the ICF(C$^{2+}$/O$^{2+}$) lies at the limit of its validity range, as $w=\mathrm{O^{2+}}/ (\mathrm{O^{+}}+\mathrm{O^{2+}}) \approx 0.96$, and the ICF is considered valid when $0.05 < w < 0.97$ \citep[see table 3 of][]{delgadoinglada14}.  On the other hand, the ICF determined by \citet{1994Kingsburgh_mnra271} for C/H in high ionisation nebulae requires knowledge of N$^{2+}$/H$^+$ that is not available here.

Following \citet{2024Gomez-Llanos_aap689}, we determine the ICF adapted to the high ionisation of the nebula using the 3MdB database\footnote{\href{https://sites.google.com/site/mexicanmillionmodels/}{https://sites.google.com/site/mexicanmillionmodels/}} \citep{2015Morisset_rmxa51}. 
The model database includes radiation- and matter-bounded nebular models produced with version 17.01 of the Cloudy code \citep{ferland17}.
We directly look for the ICF needed to produce the C/O ratio from the ionic abundances of C and O, without the intermediate steps of C/H and O/H. We use a set of 9 proxies available from the optical and UV  observations to determine the models closest to the inner region of the Necklace, namely: 
Ar$^{3+}$/(Ar$^{2+}$ + Ar$^{3+}$),
Ar$^{4+}$/(Ar$^{3+}$ + Ar$^{4+}$),
Cl$^{3+}$/(Cl$^{2+}$ + Cl$^{3+}$),
He$^{2+}$/(He$^{+}$ + He$^{2+}$),
Ne$^{4+}$/(Ne$^{2+}$ + Ne$^{4+}$),
O$^{2+}$/(O$^{+}$ + O$^{2+}$),
O$^{3+}$/(O$^{2+}$ + O$^{3+}$),
S$^{2+}$/(S$^{+}$ + S$^{2+}$), and
C$^{3+}$/(C$^{2+}$ + C$^{3+}$).

It is worth highlighting that ICF determination is based on similarities in the ionisation between the object studied and some models. The fact that multiple proxies are used (namely 9 in this work) gives more trust in the similarities, and allows the use of models that not necessary correspond in details to the object we are considering: the exact values of the ionisation parameter, the ionising SED or the optical depth (in case of matter-bounded models) are not strongly relevant, only reproducing the ionisation state matters. 

Following \citet{2009Morisset_aap507}, the distance between the $i$-th proxy and its observed counterpart is defined by: 
\[d_i = \frac{|log(M_i) - log(O_i)|}{\tau_i},\]
where $M_i$ and $O_i$ are the values of the proxy $i$ from the models and the observation, respectively, and $\tau_i$ is a normalization factor.  Using the 9 proxies and the models in the subset 3MdB\_17 under the references "PNe\_2020" and "PNe\_2021" (without any restriction), we look for models that are simultaneously within a maximal distance of 1 for each proxy, setting $\tau_i = 0.15$. We call this the look-up table method. 
The selection was carried out from the values of the table abion\_17 for both ``rad'' ionic fractions (i.e., integrated along a line of sight passing through the centre of the nebula) and ``vol'' ionic fractions (integrated throughout the entire volume of the nebula). These criteria lead to zero model that fits all the proxies together. The main issue comes from the ratio O$^{3+}$/(O$^{2+}$ + O$^{3+}$) that is incompatible with the other ones: in the same way that models are used to obtain ICFs, they can be used here to detect incoherencies between ionic fractions, namely here between  C$^{3+}$/ C$^{2+}$ and  O$^{3+}$/O$^{2+}$.
A possible solution is to compute O$^{3+}$ using a higher electron temperature, a situation that is also predicted by the 3MdB models of planetary nebulae and in some observational studies \citep[e.g.,][]{pottasch99}. We chose to set the electron temperature to 14,800 K for all the ions with IP < 50 eV, and to 20,000 K for the others (namely He$^{2+}$, O$^{3+}$, Ar$^{4+}$, and Ne$^{4+}$)\footnote{In principle, this affects the normalisation between UV and optical ranges. However, the He~{\sc ii} $\lambda1640/\lambda4686$ ratio only changes by 5\% between 15kK and 20kK, and He~{\sc ii} $\lambda1640$/H$\beta$ only by 1\%, which are well below the reported uncertainties in Table~\ref{T-COS-lines}.}.
With these criteria, we found 190 models from the "rad" set and 305 models for the "vol" set that fit all the proxies simultaneously. From these models, we extract the 3 ICFs that can be used to compute C/O from our observations, namely ICF((C$^{2+}$+C$^{3+}$)/O$^{2+}$), ICF((C$^{2+}$+C$^{3+}$)/(O$^{2+}$+O$^{3+}$), and ICF(C$^{2+}$/O$^{2+}$).

We also used another approach to compute the needed ICFs using a multi-dimensional space regressor to interpolate values from a large number of models. We selected a subset of photoionization models from 3MdB\_17 to train an Artificial Neural Network (ANN) aimed at predicting the required ICFs from specific ionic abundance ratios. The same ionic fractions as in the look-up table method described above were used as input. The training set has been obtained from 3MdB\_17 models, selecting a total of 212,001 models for “rad” and 72,571 for “vol” fitting a selection criterion of $d_i < 4$ for each of the 9 ionic fractions described above (using $\tau_i = 0.15$). This allows the regressor to be trained with relatively close models. 
80\% of each subset was used for training while the remaining 20\% was used for testing. The ANN was implemented using the scikit-learn Python library \citep{pedregosa11} and consists of three dense layers with 50, 80 and 80 Rectified Linear Unit (ReLU) neurons each. The input layer consists of the 9 proxies defined above, while the output consists of 22 ICFs, including the 3 needed for the current specific application.
Convergence was achieved using the ADAM optimization algorithm \citep{adam}. 

The resulting values for C/O are summarized in Tab.~\ref{tab:CO}, where the different methods (look-up table and ANN) have been applied to the "rad" and "vol" models from 3MdB. The errors are determined using a Monte-Carlo distribution of the line intensities corresponding to their uncertainties, and looking for the the 16th and 84th percentiles of the C/O distributions.
We checked that removing the O$^{3+}$/(O$^{2+}$ + O$^{3+}$) proxy from the inputs of both methods does not significantly change the results, thus the determination is robust against the choice of assuming $T_{\rm e}$ = 20~kK for O$^{3+}$. In Fig.~\ref{fig:CO_mosaic}, we show the individual Monte Carlo distributions of the different ICFs proposed in this work. In the case of the ICF(C$^{2+}$/O$^{2+}$), the distribution of results obtained using the approach by \citet{delgadoinglada14} is also shown.

\begin{table}[h]
    \centering
    \begin{tabular}{lcc}
        \hline
        ICF & Look up Table & ANN \\ \hline
        \multicolumn{3}{c}{Rad}\\ \hline
        ICF$\left(\frac{C^{2+} + C^{3+}}{O^{2+}}\right)$ & 0.48$+^{0.06}_{0.07}$ & 0.42$+^{0.07}_{0.06}$ \\
        ICF$\left(\frac{C^{2+} + C^{3+}}{O^{2+} + O^{3+}}\right)$ & 0.66$+^{0.07}_{0.07}$ & 0.77$+^{0.17}_{0.14}$ \\
        ICF$\left(\frac{C^{2+}}{O^{2+}}\right)$ & 0.59$+^{0.07}_{0.09}$ & 0.46$+^{0.09}_{0.08}$ \\ 
        \hline
        \multicolumn{3}{c}{Vol}\\\hline
        ICF$\left(\frac{C^{2+} + C^{3+}}{O^{2+}}\right)$ & 0.50$+^{0.06}_{0.07}$ & 0.44$+^{0.07}_{0.07}$ \\
        ICF$\left(\frac{C^{2+} + C^{3+}}{O^{2+} + O^{3+}}\right)$ & 0.69$+^{0.07}_{0.08}$ & 0.58$+^{0.12}_{0.09}$ \\
        ICF$\left(\frac{C^{2+}}{O^{2+}}\right)$ & 0.61$+^{0.07}_{0.09}$ & 0.42$+^{0.07}_{0.06}$ \\ \hline
    \end{tabular}
    \caption{C/O values using Look up Table and ANN methods.}
    \label{tab:CO}
\end{table}

Ultimately, irrespective of the methodology used to determine the C/O ratio, we reach the conclusion that the nebula is \emph{not} carbon rich.

\section{Central star properties}
\label{sec:phoebe}

Based on the analysis presented in the previous sections and in \citetalias{c11}, we can place reasonable limits on the central star's mass and temperature.  The borderline type \textsc{i/ii} abundances derived by \citetalias{c11} indicate that the progenitor is unlikely to have been particularly massive \citep[$\lesssim$2.5 M$_\odot$;][]{carigi03}, consistent with being a member of the thin disk as implied by the Gaia parallax. However, the central star is required to have become Carbon-rich, otherwise it would not have been able to chemically contaminate the companion, this places a lower limit on the initial mass, M$_i$ of 1.5--2.0 M$_\odot$ \citep{mmmb}.  This, in turn, corresponds to a final mass, M$_f$ in the range 0.55--0.6 M$_\odot$ \citep{kalirai08}.

While it is outside the scope of this paper to carry out a detailed photoionisation modelling of the nebula, we did undertake some simple modelling runs to try and constrain the temperature of the central star using \textsc{cloudy} \citep{ferland17}.  The main issue is that the [O~{\sc i}]$\lambda$ 6300 emission line is not detected in the inner region, pointing to a matter-bounded nebula (as would be expected from the very low ionised mass of the nebula). This leads to a strong degeneracy between: 
\begin{itemize}
\item The central source effective temperature
\item The ionisation parameter, U, which combines the luminosity of the central source, the distance of the nebula from this central source, the electron density and an eventual filling factor
\item The optical depth of the region to the Lyman continuum photons (less than unity in the case of matter-bounded nebulae) 
\end{itemize}
We found that no satisfactory solution can be found for effective temperatures of the ionising source outside of the 140--170~kK domain. Between these two values, the main emission line intensities can be reproduced with an acceptable tolerance. The solution is still not totally presentable, as the electron temperature derived for these models is significantly higher than the observed value derived by \citetalias{c11}.

This temperature (140--170 kK) is roughly the maximum temperature reached by the solar metallicity ($Z_0=0.02$) models of \citet{mmmb} for M$_i$=1.5 M$_\odot$ and M$_i$=2.0 M$_\odot$ (M$_f$=0.576 M$_\odot$ and M$_f$=0.580 M$_\odot$, respectively).  Furthermore, the kinematical age of the nebula \citepalias[$\sim$5.4 kyr;][]{c11} is roughly consistent with the crossing time of the M$_i$=1.5 M$_\odot$ model (4.49 kyr), while the M$_i$=2.0 M$_\odot$ model is a factor of two shorter (although the starting point for post-AGB ages is notoriously arbitrary).  Estimates of the mass of the nebula are extremely low, which one could perhaps consider as further support for a lower mass progenitor, however the sum of the ionised and molecular mass \citep[$<0.01$ M$_\odot$;][]{santander22} is irreconcilably low even for a very low mass progenitor. It is important to note that higher mass models (3--4~M$_\odot$), which would not be as consistent with abundance pattern of the nebula but which would still become C-rich, pass through the inferred temperature regime twice. Once as they cross the HR diagram, at unfeasibly early times (tens to hundreds years). The second time as they cool at much lower (orders of magnitude lower) luminosities (in fact, the 3~M$_\odot$ model returns to this temperature at an age of only 1000 years, too young to be consistent with the kinematical age).

Taking a central star mass of 0.58 M$_\odot$, the radial velocity semi-amplitude for the secondary, $K_2$, of roughly 80 \kms{} implies a mass for the companion of $\leq$0.8 M$_\odot$. The very high amplitude of reflection effect is indicative of a physically large companion \citep{jones22}, potentially consistent with an earlier spectral type than most main sequence companions to central stars of planetary nebulae \citep[which are typically late K- to early M-type;][]{jones15}.  Assuming that the contribution to the spectral energy distribution from the central star is roughly flat ($g-r \approx r-i \approx 0$), a relatively large and massive companion would be consistent with the colours of the central star system around minimum $g-r \approx 0.65$ and $r-i \approx 0.4$, or $g-r \approx 0.25$ and $r-i \approx 0.13$ after dereddening, corresponding to a mid-F-type companion \citep{covey07}.  However, for an inclination of 59$^\circ$ (as one would expect from the nebular inclination), an appreciable fraction of the irradiated hemisphere of the companion would still be visible at inferior conjunction (just as indicated by the detection of irradiated lines close to inferior conjunction in Sec.\ \ref{sec:rvs}) and thus, even at photometric minimum, the colours of the companion may be shifted bluer (and thus towards earlier spectral types) by the irradiation from the central star.  Furthermore, the lines from which the RVs of the companion are measured are shifted from the centre of mass of the companion, originating only from the irradiated hemisphere, thus the mass of the companion would be appreciably lower than the $\sim$0.8~M$_\odot$ upper-limit derived above.

As the central star of the Necklace is not eclipsing, it is difficult to place strict constraints on the stellar temperatures and radii.  However, a reasonable fit to the light and RV curves could be obtained by fixing the orbital inclination to that of the nebula and the parameters of the primary to values obtained from the evolutionary tracks of \citet{mmmb} at the kinematical age of the nebula, as discussed previously.  The spectrum and limb-darkening of the primary was approximated using T\"ubingen Model Atmosphere Package \citep[TMAP;][]{rauch03,werner03,reindl23} as implemented in the \textsc{phoebe2} code \citep[Jones in prep.]{phoebe2,phoebe4,phoebe5}.  The parameters of the secondary (mass, temperature, radius and bolometric albedo) were allowed to vary freely, with those providing the best fit listed in Table\ \ref{T-binmodel}. The radial velocities of the secondary, as measured by the irradiated lines, were assumed to be well-represented by the centre-of-light radial velocities (rather than centre-of-mass) in the \textsc{phoebe2} model which, due to the high levels of irradiation are strongly displaced towards the inner Langrange point of the system.

Note that, due to the various assumptions in the model, we do not quote the statistical uncertainties on the derived parameters as they are not representative of the true uncertainties. Furthermore, given the large uncertainties on the distance and extinction, the fitting was not performed in absolute units, meaning that the likelihood function accounted for the amplitudes of variability but not the true observed magnitudes.  Nevertheless, the fit reasonably reproduces the observed magnitudes for a distance of $\sim$4.6~kpc (entirely consistent with statistical distance estimates and the Gaia parallax) and a visual extinction of $\sim$2 magnitudes -- appreciably larger than that derived by \citetalias{c11} ($A_V\sim1.2$~mag), but comparable to the value derived by \citet[$A_V\sim2.4$~mag]{dharmawardena21}. Applying the method of \citet{csukai25}, which fits the central star extinction based on catalog photometry, leads to a similarly high extinction of $A_V\sim2.7$~mag based on five SED points in the optical (redder points are excluded by their algorithm to the observed excess, see Sec.\ \ref{sec:discussion}).  The uncertainties on the derived extinction are large\footnote{The central star of the Necklace was excluded from the sample of \citet{csukai25} due to the relatively compact nebula but would also not have been considered a good fit according to their criteria, likely as a result of the impact of the companion and the intrinsic variability due to irradiation on the observed spectral energy distribution.} but the derived value is certainly consistent with a larger extinction for the central star than has been previously measured for the nebula. One may expect such an increased extinction should there be an appreciable amount of circumbinary material remaining following the ejection of the CE.  Alternatively, if the remnant mass were larger than assumed here, the model luminosity would be lowered potentially to a level consistent with the nebular extinction. However, this does not seem particularly likely as much more massive remnants (which reach the correct temperature at approximately the right post-AGB age as they cool) would have luminosities that are an order of magnitude lower, while slightly more massive remnants reach the correct temperature at much young ages than implied by the kinematical age of the nebula.

Ultimately, while the \textsc{phoebe2} fit is not perfect (as mentioned in Sec.\ \ref{sec:ephem}, the statistical uncertainties of some of the photometric observations appear to be underestimated) and it is based on a number of assumptions, we can draw some interesting conclusions:

\begin{itemize}
\item A satisfactory fit to the light and RV curves is found using central star parameters which lie on single star evolutionary tracks, while many other post-CE central stars seem to be inconsistent with these tracks \citep{jones19}. If this best-fit set of parameters is indeed representative of the system then it might suggest an interesting scenario in which the CE occurred so close to the end of the central star's AGB evolution that it did not have a significant effect on its post-AGB evolution, perhaps consistent with the CE having occurred at the very tip of the AGB.  This would be expected for a 1.5 M$_\odot$ progenitor (as assumed here), and would also be consistent with the remnant having become Carbon-rich similarly late in its evolution.
\item The current mass ratio, $q=M_2/M_1$, is approximately unity (or lower).  As such, the pre-CE mass ratio must have been much lower, consistent with the idea that extreme mass ratios are a requirement for unstable mass transfer and thus for a CE \citep[and references therein]{jones20ce}.
\item The secondary is significantly inflated (and hotter) than would be expected for its mass as has been found in several other post-CE binary central stars \citep{jones15}.  This is thought to be a consequence of mass transfer \citep[which clearly occurred here as evidenced by the carbon contamination of the companion;][]{miszalski13}.
\item The extinction of the central star may be larger than that derived for the nebula by \citetalias{c11}.
\end{itemize}

\begin{table}
\caption{Binary stellar parameters from \textsc{phoebe2} modelling. Parameters marked with an asterisk were fixed.}
\centering
\begin{tabular}{rl}       
\hline\hline\\                    
Parameter & Value  \\
\hline\\[-7pt]              
Primary mass$^*$, $M_1$ (M$_\odot$) & 0.576 \\
Primary effective temperature$^*$, $T_1$ (kK) & 148 \\
Primary radius$^*$, $R_1$ (R$_\odot$) & 0.085 \\
Secondary mass, $M_2$ (M$_\odot$) & 0.55 \\
Secondary effective temperature, $T_2$ (kK) & 5.6  \\
Secondary radius, $R_2$ (R$_\odot$) & 0.99  \\
Secondary bolometric albedo, $A_2$ & 0.7\\
Binary inclination$^*$, $i$ ($^\circ$) & 59\\
\hline\\
\end{tabular}
\label{T-binmodel}
\end{table}

\section{Discussion}
\label{sec:discussion}

Analysis and modelling of the light and RV curves of the central binary star of the Necklace reveal the system to comprise a hot post-AGB star, with parameters consistent with an initial mass of 1.5--2.0 M$_\odot$ close to the limit to become carbon-rich \citep{marigo17,abia20}, along with a highly inflated dC star companion.  This is consistent with the previous studies which suggested that the dC companion became chemically contaminated during a period of pre-CE accretion which also led to the formation of the nebula's bipolar jets \citep{miszalski13}.

Analysis of the physical and chemical properties of the Necklace indicates that the nebula is likely \emph{not} carbon-rich (although the C/O ratio is rather uncertain).  This is difficult to reconcile with presence of a dC companion to the central star, as the nebular progenitor would thus have been expected to become carbon-rich before both contaminating the companion and forming the nebula.

\begin{figure}
\centering
\includegraphics[width=\columnwidth]{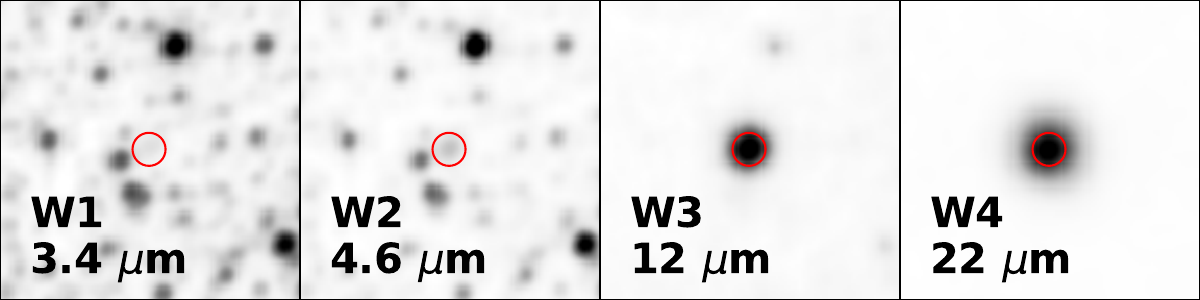}
\caption{
WISE imagery of the Necklace highlighting the clear excess in both bands 3 and 4, possibly attributable to carbonaceous dust, although the estimated dust mass based on IRAS photometry is on the order of $1\times10^{-5}$~M$_\odot$ only.  North is up, East is left, and all images measure 180''$\times$180'' and the location of the Necklace is marked with a red circle.}
 \label{F-SED}
\end{figure}

One potential explanation is that while the ionised material is carbon-poor, a significant fraction of the carbonaceous material could be in the form of dust \citep{toala21}.  This is particularly attractive when one takes into account the low mass of the ionised nebula \citepalias{c11}.  Wide-field Infrared Survey \citep[WISE;][]{wise} imagery indicates a strong excess in both bands 3 (12\micron{}) and 4 (22\micron{}), confirmed by the IRAS 25\micron{} and 60\micron{} fluxes \citep[IRAS provides only upper limits for the 12\micron{} and 100\micron{} fluxes;][]{iras}, which could be consistent with a large (carbonaceous) dust mass (see Fig.~\ref{F-SED}).  Similarly, the large extinction implied by the best fitting binary model, $A_V\sim2$ mag c.f. $A_V\sim1.2$ mag from nebular spectroscopy, could be consistent with an appreciable amount of circumbinary dust.  Ultimately, only infrared spectroscopy and detailed dust modelling can probe the validity of this hypothesis, but one can use the IRAS continuum fluxes at 25\micron{} and 60\micron{} to estimate the dust temperature and mass \citep[equations 2 and 4]{muthumariappan20}. This leads to a dust temperature of 120~K and dust mass of $\sim1.1\times10^{-5}$~M$_\odot$ (for a distance of 5.4~kpc), both fairly typical for a PN and in no way indicative of a hidden C-rich mass of dust (although the dust mass is highly dependent on the distance which is rather uncertain).

It is important to note that the analysis presented in section \ref{sec:co} does depend on the extinction for the normalisation of the UV and optical spectra, with the uncertainties there propagating through to the measured abundances.  However, we repeated the ANN and look-up table analyses assuming a much higher extinction $c(\mathrm{H}\beta)$=1.1 (equivalent to an $A_V=2.2$ mag) and find that the while the C/O ratio does change it is still below unity in all cases.  As such, even if the nebular extinction is underestimated (i.e. the increased extinction is not only circumbinary), the conclusions here remain unchanged.

An alternative explanation for the apparent incongruence between nebular and stellar abundances would be that the nebula is chemically very inhomogeneous, with the region covered by the HST-COS aperture being particularly carbon-poor while other regions (the knots, for example) would be carbon-rich.  Strong chemical inhomogeneities have been observed in a number of post-CE PNe \citep[e.g.;][]{wesson18}, however the lack of observed recombination lines in the optical spectrum of the Necklace would seem to indicate that its abundance discrepancy factor is relatively low \citep[\citetalias{c11},][]{corradi15}.  Additional pointings of UV spectroscopy, for example centred on a bright knot complex, or deep integral field spectroscopy could potentially reveal any as yet unobserved chemical inhomogeneities.

A third, perhaps less likely, possibility is that the majority of the nebular material is formed from material lost while the central star was still oxygen-rich, while the companion was contaminated only at the very end of the AGB once the star had become carbon-rich.  This would require very significant fine tuning and/or a very late thermal pulse \citep[VLTP;][]{mmmb}, although this might be consistent with the initial mass of the central star being so close to the limit to become carbon-rich (and thus would only become carbon-rich during the last or last few thermal pulses and, indeed, the central star parameters used in the \textsc{phoebe2} modelling come from a VLTP model).  Furthermore, the maximum RGB and AGB radii for a 1.5 M$_\odot$ progenitor are relatively similar, thus the CE event must have happened close to the end of the AGB in order to have avoided the CE while on the RGB \citep{jones20ce}, potentially making this hypothesis slightly more plausible.

Ultimately, The Necklace continues to be a highly enigmatic object: the only post-CE PN known to host a dC star and the most extreme example of the so-called nebular \emph{missing mass problem} \citep{santander22}.  While we find the nebula to be O-rich, it seems clear that the amount of C-rich material accreted by the companion must have been significant, with perhaps as much as a few tenths of a solar mass (a very significant fraction of its current mass) being required to make it C-rich for all reasonable donor C/O ratios \citep{miszalski13}.  This material was likely accreted during a period of wind Roche lobe overflow that also led to the formation of the bipolar jets, prior to the runaway Roche lobe overflow which led to the CE event and the formation of the Necklace's equatorial ring.  However, further study is required to fully understand the evolution of the Necklace and its central stars.

\begin{acknowledgements}

We thank the anonymous referee for their comments which improved the content and clarity of this paper.
DJ, JG-R and RLMC acknowledge support from the Agencia Estatal de Investigaci\'on del Ministerio de Ciencia, Innovaci\'on y Universidades (MCIU/AEI) under grant ``Nebulosas planetarias como clave para comprender la evoluci\'on de estrellas binarias'' and the European Regional Development Fund (ERDF) with reference PID2022-136653NA-I00 (DOI:10.13039/501100011033). DJ and PS also acknowledge support from the Agencia Estatal de Investigaci\'on del Ministerio de Ciencia, Innovaci\'on y Universidades (MCIU/AEI) under grant ``Revolucionando el conocimiento de la evoluci\'on de estrellas poco masivas'' and the the European Union NextGenerationEU/PRTR with reference CNS2023-143910 (DOI:10.13039/501100011033).  GAGP and LS acknowledge support from UNAM PAPIIT Grants IN107625  and IG101223.  CM acknowledges support from grant UNAM / PAPIIT - IN101220 and from a visitor grant to the IAC under Severo Ochoa excellence program CEX2019-000920-S. PR-G acknowledges support by the Agencia Estatal de Investigaci\'on del Ministerio de Ciencia e Innovaci\'on (MCIN/AEI) and the European Regional Development Fund (ERDF) under grant PID2021--124879NB--I00. MSG acknowledges support from the I+D+i PID2023-146056NB-C21 (CRISPNESS/MESON), funded by the AEI (10.13039/501100011033) of the Spanish MICIU and the European Regional Development Fund (ERDF) of the EU. MMRD acknowledges support by grant PID2022-137779OB-C41, funded by the Spanish Ministry of Science, Innovation and Universities/State Agency of Research MICIU/AEI/10.13039/501100011033.

This article is based on observations made at the Observatorios de Canarias del IAC with the the 2.5-m~Isaac Newton (INT) and 4.2-m~William Herschel (WHT) telescopes operated by the Isaac Newton Group of Telescopes; the 2.0-m~LT operated by Liverpool John Moores University; the 2.56~m Nordic Optical Telescope owned in collaboration by the University of Turku and Aarhus University, and operated jointly by Aarhus University, the University of Turku and the University of Oslo, representing Denmark, Finland and Norway, the University of Iceland and Stockholm University; and the 10.4~m Gran Telescopio Canarias (GTC), all at Observatorio del Roque de los Muchachos on the island of La Palma.  Also based on observations obtained with the 0.8~m IAC80 Telescope owned and operated by the IAC at the Observatorio del Teide on the island of Tenerife.  Based on observations obtained with the Apache Point Observatory 3.5-meter telescope, which is owned and operated by the Astrophysical Research Consortium. The data presented here were obtained in part with ALFOSC, which is provided by the Instituto de Astrof\'isica de Andalucia (IAA) under a joint agreement with the University of Copenhagen and NOT. Also based on observations made with the NASA/ESA Hubble Space Telescope, obtained at the Space Telescope Science Institute, which is operated by the Association of Universities for Research in Astronomy, Inc., under NASA contract NAS 5-26555. These observations are associated with program 13424 and the Hubble Heritage Project. This publication makes use of data products from the Wide-field Infrared Survey Explorer, which is a joint project of the University of California, Los Angeles, and the Jet Propulsion Laboratory/California Institute of Technology, funded by the National Aeronautics and Space Administration. This research made use of Astropy,\footnote{http://www.astropy.org} a community-developed core Python package for Astronomy \citep{astropy:2013, astropy:2018}.

\end{acknowledgements}

\bibliographystyle{aa} 
\bibliography{necklace}

\appendix
\section{Individual distributions of the C/O ratio}
\label{app:c_o_dist}
\begin{figure}[!h]
\centering
\includegraphics[width=\textwidth]{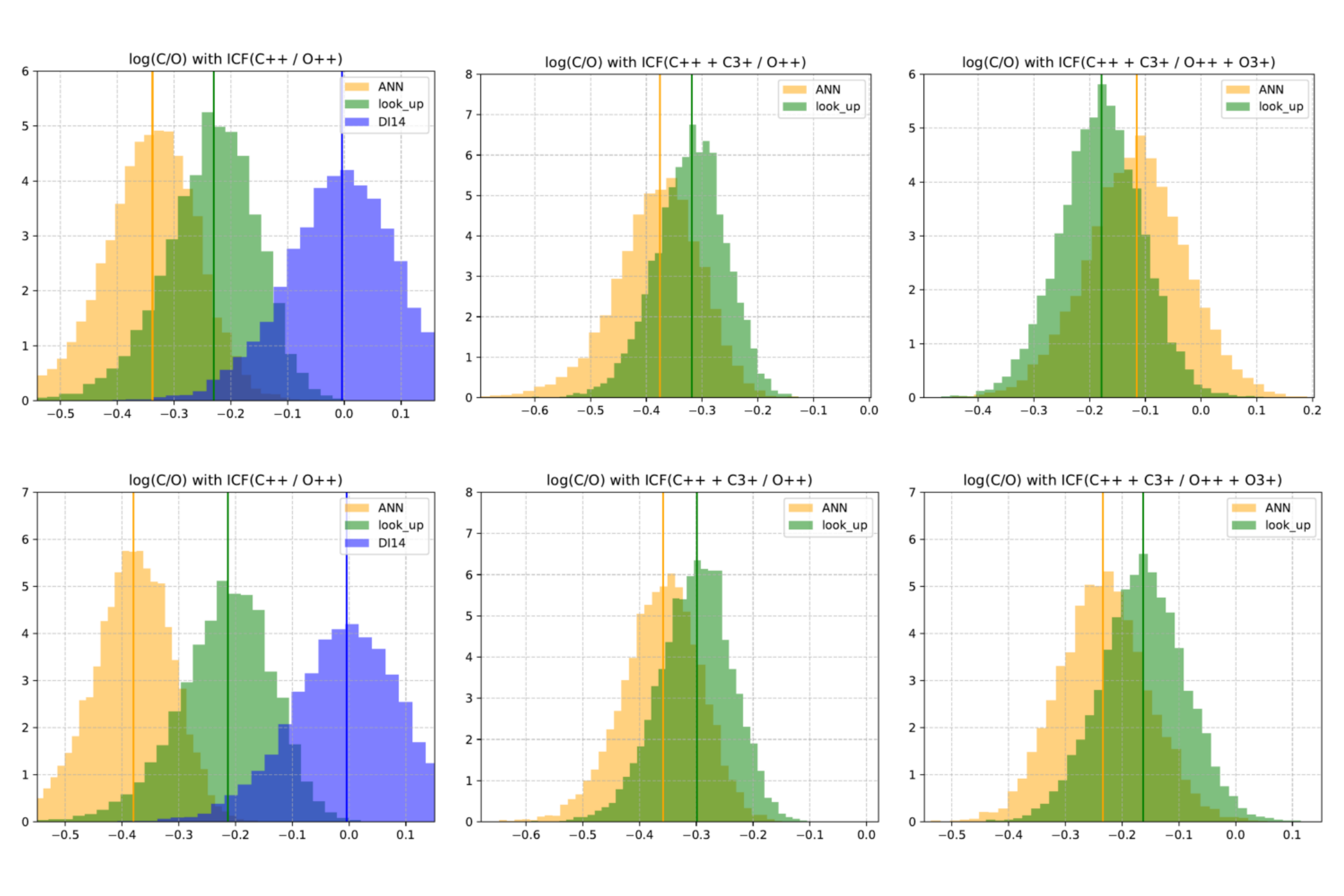}
\caption{Monte Carlo distributions of the C/O ratio using the three ICF methods: the ICF by \citet{delgadoinglada14}} (purple), the Look-up method (green), and ANN (orange).The first row corresponds to the models labeled as “rad,” and the second to those labeled as “vol”. The vertical lines indicate the median of each distribution.
 \label{fig:CO_mosaic}
\end{figure}

\end{document}